\magnification 1200
\tolerance 1500
\def \cl {\centerline}
\def \y {\'{\i}}
\def \noi {\noindent}
\def \bs {\bigskip}
\def \ms {\medskip}
\def \gc {Garc\'{\i}a-Col\'{\i}n}
\def \sm {\smallskip}
\def \grad{\hbox{\rm grad}~}
\def \ub {\underbar}
\baselineskip 18pt
\cl {\bf Some Thoughts about Non-equilibrium Temperature}
\vskip .5cm  \cl{by}
\vskip .5cm
\cl {L.S. \gc ; Departamento de F\'\i sica,}
\cl {Universidad Aut\'onoma Metropolitana-Iztapalapa,}
\cl {Apdo. Postal 55-534, 09340 M\'exico D.F.}
\vskip .5cm  \cl{and}
\vskip .5cm
\cl{V. Micenmacher; Instituto de F\y sica,}
\cl{Facultad de Ingenier\y a, Universidad de la Rep\'ublica,}
\cl{C.C. 30, Montevideo, Uruguay.}
\vskip .5cm
\cl{Abstract}
\sm
The main objective of this paper is to show that, within the present
framework of the kinetic theoretical approach to irreversible
thermodynamics, there is no evidence that provides
a basis to modify the ordinary Fourier equation relating the heat flux
in a non-equilibrium steady state to the gradient of the local
equilibrium temperature.
This fact is supported, among other arguments, through the kinetic
foundations of generalized hydrodynamics. 
Some attempts have been recently proposed 
asserting that, in the presence of non-linearities of the state
variables, such a temperature should be replaced by the non-equilibrium
temperature as defined in Extended Irreversible Thermodynamics. In the
approximations used for such a temperature there is so far no evidence
that sustains this proposal. 

\bs
\noi PACS: 05.70.Ln, 05.20.Dd
\vfill\eject
\noi {\bf I. Basic Concepts.}
\sm
The problem of extending uniquely defined thermodynamic quantities for
equilibrium states of many body systems to non-equilibrium states
remains still highly controversial. This situation is particularly
striking when referring to the concepts of entropy and temperature. For
a one-component fluid in equilibrium, the three basic laws of
thermostatics are condensed in the equation
$$
TdS=dU+pdV,\eqno (1)
$$
where all symbols have their ordinary meaning. Thus, the equilibrium
temperature $T$ is related to the two state variables $S$, the entropy
as defined by Clausius, and $U$, the internal energy, by the relation
$$
{1\over T}=\left({\partial S\over \partial U}\right)_V.\eqno (2)
$$
Eq. (2) is easily generalizable to equilibrium states of more
complicated systems. Any attempt to extend eq. (2) to a
non-equilibrium state of any arbitrary system would imply that not only
$U$ but also the entropy $S$ is uniquely defined for such a state. This
requirement is met in Linear Irreversible Thermodynamics (LIT) through
the local equilibrium assumption${}^{1,2}$ whereby in (2) all
quantities depend on the position $\vec r$ of an
infinitesimal element of the system at time $t$. The local equilibrium
temperature is thus a quantity which uniquely characterizes a system's
element whose characteristic length  is much longer that a mean
free path, but much smaller than the lenght of the
container holding the system. Moreover, eq. (2) contrary to statements
often issued in the literature, is not the definition of temperature.
It provides an equation of state of the intensive parameter $T$ which,
after use is made of the basic contents of the zeroth law, is shown to
posses all properties required to define what we invariably understand
as equilibrium temperature${}^{3}$.
\ms
Beyond local equilibrium the question
which has caught the attention of many workers in this field concerns
the validity of an appropriate extension of eq. (2).
This question as well as how to formulate a theory for irreversible
processes within the context of the laws of thermodynamics has been
exhaustively discussed in other papers${}^{4a, b}$. Within this
context, it is pertinent to mention that many
efforts have been made in the past to define the entropy of a
non-equilibrium state of an arbitrary system, none of which has led to a
clear cut answer. In fact, twenty years ago in a not very well known
paper, J. Meixner${}^{5}$, one of the founders of LIT, gave very
convincing arguments to show that it is very unlikely that a
non-equilibrium state function playing the role of the entropy may be
uniquely defined. Indeed, summarizing his ideas one may assert that
the conclusion reached by him is that such a function, either cannot be
defined, or if it can, then it may be done so in an infinite number of
ways. Without being so drastic in this last assertion, a similar
conclusion was held by Grad over thirty years ago${}^{6}$. In
exploring recent literature on this question these conjectures seem to
hold true in a more restricted sense. Taking a closed (constant
mass) thermodynamic system undergoing an irreversible process, and
assuming that the pertinent state variables describing the system's
states during such a process are stochastic variables, one can provide
several definitions for non-equilibrium entropies when the process is
Markovian${}^{7}$. Yet if the system's state at a certain time $t$
is still characterized by the histories of the independent state
variables up to such a time, the process is non-Markovian, and then an
entropy-like function has resisted a definition . These, in different
words, are the ideas sustained by Meixner. Thus, according to this
conjecture one would have none or as many intensive parameters
formally playing the role of $1/T$ in eq. (2) as entropy-like
functions one could define. Nevertheless, the only possible definition
of non-equilibrium temperature consistent with the laws of
thermodynamics is already implicit in Clausius inequality${}^{4b}$.
\ms
In view of this result, which holds true for non-equilibrium 
states, one may raise the question about the physical
meaning of the quantity $\Theta$ defined by:
$${1\over \Theta}=\left({\partial\eta\over\partial u}\right)_{v,\dots},\eqno 
(3)$$
where $\eta$ is one of such entropy-like functions and $u$ the specific 
internal energy, a dynamical quantity which may be uniquely defined. Here 
$(~~)_{v,\dots}$ means that in eq. (3) the specific volume and other 
pertinent extensive state variables are kept fixed in taking the 
partial derivative. However, following the postulates of Extended
Irreversible Thermodynamics (EIT) we shall
refer to $\Theta$ as a non-equilibrium temperature${}^{8}$
and to $\eta$ as an entropy-like function. As far as
$\Theta$ is concerned, here taken as an unknown function of all the
variables on which $\eta$ depends, it has been proposed to
regard it as the phenomenological temperature, that is as the reading
provided by a thermometer introduced into the system assuming that
local thermal equilibrium is reached by both systems. What we wish to
emphasize in this paper is that so far, available molecular
theories have not provided a suitable and reliable molecular
interpretation for this quantity, whereas they reinforce the unique
validity of the local equilibrium temperature.
\ms
In Section II we will briefly summarize the arguments that motivated
this work. In Section III we go into the core of the paper showing that
the kinetic theoretical considerations are inssufficient to give a
molecular interpretation of $\Theta$. We also argue that at present,
other microscopic methods have not yet solved the question and in our
opinion will hardly do so.
\vskip .75cm
\noi {\bf II. Background.}
\sm
For the sake of simplicity and to make contact with other approaches
we shall restrict ourselves to the examination of a system defined by a
rigid heat conductor${}^{8}$ characterized by its internal energy density 
$u(\vec r,t)$ and the heat flux $\vec q(\vec r ,t)$ raised to the
status of independent variable. According to the basic postulate of
M\"uller ${}^{8c,9}$ concerning the phenomenological structure of
EIT, we assume that a regular and continuous function exists depending
on the independent state variables, which acts like a generalized
thermodynamic potential namely, an entropy like function. In the case
of our rigid conductor we call this function, as in eq. (3),
$\eta=\eta(u,\vec q)$, and therefore,
$$
d\eta=\left({\partial\eta\over \partial u}\right)_{\vec q}du+
\left({\partial\eta\over \partial \vec q}\right)_u{\bf\cdot}d\vec q
\eqno (4)$$
By analogy with eq. (1) and according to eq. (3) we define the
non-equilibrium temperature $\Theta(u,\vec q)^{-1}$.
Notice that $\eta$ and $\Theta$ being both scalars and functions of
$\vec q$, must be functions of all scalar invariants of $\vec q$, namely
$q^2=\vec q{\bf\cdot}\vec q$. Thus, to the lowest order in $\vec q$, we may
write that 
$$\eqalignno{
\eta=&s_0-{\alpha\over 2}q^2,&(5)\cr  {1\over\Theta}=&{1\over T}-
{\alpha'\over 2}q^2.&(6)\cr}
$$
where $s_o(u)$ is the local equilibrium entropy, $\alpha$ depends on
$u$ and $\alpha'=d\alpha/du$, and $T$ is the local equilibrium temperature
as related to the local state variable $u$ 
through $s_0$. Emphasis should be made however, that in eq.
(6) the quantity $\Theta$ is \ub{not} the full non-equilibrium
temperature as defined in eq. (3), but only an approximate expression
that contains deviations from $T^{-1}$ by quadratic terms in $\vec q$.
\ms
On the other hand $(\partial\eta/\partial\vec q)_u$ being a vector,
must in general be of the form
$$
\left({\partial\eta\over \partial \vec q}\right)_u=-A(u,\vec q)\vec q.
$$
and using eq. (5) one gets $A(u,0)=\alpha(u)$. The time derivative of $\eta$
is then written
$$
\dot{\eta}=\Theta^{-1}\dot{u}-A\vec q{\bf\cdot}\dot{\vec q}.
\eqno (7)$$
where $\dot {x}\equiv dx/dt$.
\sm
Nevertheless, eq. (6) is the starting point of recent discussions${}^{8,10}$
concerning the possibility of measuring $\Theta$ by some kind of
experiments. The argument invoked in these efforts is that eq. (7),
the equation for the energy balance and the definition of a vector
$\vec J_\eta=\Theta^{-1}\vec q$, leads to a balance
equation for $\eta$, in which the production term $\sigma_\eta$ has the
form
$$
\sigma_\eta=\vec q(\nabla\Theta^{-1}-\rho A\dot{\vec q}),\eqno (8)
$$
and this quantity, arguing compatibility with the second law, is
assumed to be non-negative. Then it is proposed that due to the
structure of eq. (8)
$$
\nabla\Theta^{-1}-\rho A\dot{\vec q}=\mu\vec q\eqno (9)
$$
and $\mu\geq 0$. Therefore the time evolution equation for $\vec q$ is
to be of the Maxwell-Cattaneo-Vernotte (MCV) form, namely
$$
\tau\dot{\vec q}+\vec q=-\lambda\nabla\Theta,\eqno (10)
$$
where $\tau=\rho A/\mu$ and $\lambda=1/\mu\Theta^2$. Now, $\tau$, $\lambda$
and $\Theta$ are in 
general functions of $u$ and $\vec q$,so that eq. (10) is a highly
nonlinear equation for $\vec q$. When we limit ourselves to the
approximation used in eq. (5), $\lambda$ and $\tau$ become functions of
$u$ only. Moreover, the complete linearization of eq. (10) is achieved
with $T$ in place of $\Theta$, yielding the ordinary MCV equation${}^{8c}$.
\ms
The quantity $\Theta$ may be pressumably 
identified with the phenomenological temperature, instead of $T$ as
occurs in LIT. Thus, when we consider 
steady state situations, eq. (10) acts as a generalized Fourier law, namely
$$
\vec q=-\lambda\nabla\Theta,\eqno (11)
$$
that may give some physical content to the non-equilibrium temperature
defined in eq. (3). Once we accept that the relation between
temperature and heat flux is given by eq. (10) or (11), then $\Theta$
becomes the only temperature that can be measured${}^{10b}$, and it is
not possible to measure the local equilibrium temperature $T$
independently, but only to calculate it through eq. (6).
\ms
The question we want to raise in this paper concerns the possibility
of sustaining the validity
of eq. (10) from both, kinetic and statistical
mechanical models of matter. As we shall argue in the following
section there is no such evidence. That means that
to the order of approximation involved in eqs. (5) and (6),
the difference between $T(\vec r,t)$ and $\Theta(\vec r,t)$, if any, is
completely negligible. In all cases the quantity that preserves its
physical identity is $T$. This is shown to be true in general in ref. (4b).
\bs
\noi {\bf III. Kinetic and Statistical Interpretation of the
Non-equilibrium  Temperature.}
\sm
We begin by considering the definition of the
non-equilibrium temperature $\Theta$ in the case of a dilute gas whose
dynamics is governed by the Boltzmann equation${}^{11}$. We recall the
reader that when this equation is multiplied by $\ln f$, (we assume that
the integrals converge and that $f$ vanishes in the boundaries of the
velocity subspace) integration over the $\vec v$-space leads to a
balance type equation for the entropy as defined by 
$\rho\eta=-k\int f \ln fd\vec v$,
in which the entropy production is non-negative. This property is
inherent to the Boltzmann equation and must be obeyed by any exact
solution $f$ of the full non-linear equation. Such solutions are still
unknown for realistic systems. One therefore resorts to approximate
methods of solution among which Grad's moments solution${}^{11}$ is
rather pertinent to the problem posed here. This method is based upon
the expansion of the single particle distribution function around its
local equilibrium (Maxwellian) form, as an infinite series in Hermite
tensorial polynomials whose coefficients $a^{(s)}(\vec r,t)$, known as
Grad's moments, perform here the role of the local state variables
describing states beyond the local-equilibrium one. Nevertheless, for
practical reasons the infinite series is in general truncated at some
stage, although there exists no well defined criterion that indicates
how or where such truncation must be performed. In the case of the
thirteen moment approximation${}^{11,12}$, it is well known that the
physical fluxes namely, the stress tensor $\cal T$ and the heat flux
$\vec q$ are raised to the status of independent variables. Taking a
system such that its states depend only on $\vec q$ (shear free
systems), then the single particle distribution function is given by
$$
f(\vec r,\vec c,t)=f^{(0)}(\vec r,\vec c,t)\left\{1+{2\over 5}{m\over
pk_{{}_B}T}\left({mc^2\over 2k_{{}_B}T}-{5\over 2}\right)\vec
q\cdot\vec c\right\}\quad .\eqno (12)
$$
Here $f^{(0)}$ is the ordinary local Maxwellian distribution function, $m$
the mass of the particles $p$ and $T$ the local pressure and
temperature respectively, $k_{{}_B}$ Boltzmann's constant, $\vec q$
the heat flux and $\vec c\equiv\vec v-\vec u(\vec r, t)$ the chaotic or
thermal velocity. $\vec u(\vec r,t)$ is the local hydrodynamic 
velocity defined as the first moment of $f$. It is important to recall
that in kinetic theory of dilute gases the local equilibrium
temperature is interpreted as $\langle{1\over2}mc^2\rangle$ where the
averages must be taken with (12). Thus, there is no room to bring in
the non-equilibrium temperature $\Theta$ in any logical way.
Nevertheless, when the
entropy of the gas is computed using eq. (12) one performs an
integration in which $\ln f=\ln f^{(0)}+\ln(1+x)$, where $1+x$ is the
expression within the curly brackets in (12) and $x$ is regarded as
small compared with 1. So that retaining terms up to order $x^2$, one is
lead to the result${}^{13}$
$$
\rho\eta=\rho s_0-{m\over 5pk_{{}_B}T^2}\vec q\cdot\vec q~~,\eqno (13)
$$
which is precisely of the form expressed by eq. (5). The coefficient
$\alpha=m/5pk_{{}_B}T^2$ can be rewritten in terms of $v$ and
$u$, using the local forms of the equation of state and the equation
for the internal energy. Thus, eq. (13) leads to a temperature
$\Theta$ whose form is that of eq. (6) and need not be written
explicitely. This result shows an inconsistency between the 
tenets of kinetic theory of gases where $T(\vec r,t)$ is defined as
$\langle 1/2 mc^2\rangle$ and eq. 6. In the best case, it simply 
relates the local equilibrium temperature $T(\vec r,t)$ with $\Theta$ 
but we insist, defines neither of them.
\ms
Now, according to our discussion in section II, eq. (13) should be
consistent with an equation of motion for $\vec q$ of the form exhibited
by eq. (10). We must then examine the nature of the equation for $\vec
q$. The full non-linear equation is complicated enough but certainly
not of the form required by (10)${}^{12,13}$. Its linearized
version, the one consistent with eq. (13), is
$$
{\partial\vec q\over \partial t}+{5pk_{{}_B}\over 2m}\grad T=-{16\over
15}n\Omega^{(2,2)}\vec q~~,\eqno (14)
$$
where $\Omega^{(2,2)}$ is a well known collision integral. Notice that
eq. (14), as expected from the interpretation of $T(\vec r,t)$ contains
grad $T(\vec r,t)$ and not grad $\Theta$ with $\Theta$ given by eq
(6). The non-linearities in $\eta$ arising from the quadratic term 
in the heat flux do not have any influence whatsoever upon the kinetic
interpretation of $T(\vec r,t)$. In the best of cases, this simply 
means that to the order of
approximation involved in eqs. (5) and (6) the difference between 
$T$ and $\Theta$ is negligible. In fact the only physically meaningful
quantity is $T$. 
\ms
One may pursue this analysis to incorporate more and more
moments into the description of the states of the gas. In the case of
26 moments${}^{15}$, which is illustrative enough, one gets, using
the same approximations as above, linearizing the $\ln f$ term to
compute the entropy and the equations of motion for the moments,
that $\rho\eta$ may always be written as the difference of two terms,
$\rho s_0$ minus a sum of terms, \ub{some} quadratic in the fluxes.
Nevertheless, for the same reason as above the difference between the
two temperatures, remains negligible. This result should not 
surprise us. If Grad's, or any other method is devised to seek
approximate solutions to the Boltzmann equation, the only well
defined temperature is $T$ taken as the average of the thermal
kinetic energy. The quantity $\Theta$ as calculated from eq. (3)
has no physical meaning as it depends on the degree of
approximation used to compute the function $\eta$. 
Thus, the kinetic theory of a dilute gas where no
potential energy contributes to the total energy, manifestly
indicates that the only variable which is justified as a physically
meaningful temperature, is the local equilibrium temperature
$T(\vec r,t)$. This is also consistent with the laws of 
thermodynamics${}^{4c}$.
 \ms
Further support for the above conclusions comes from the full solution to 
the Boltzmann equation.
Indeed, the result arising from the full
infinite series for $f$ namely, when \ub{no} truncations
are performed has been recently obtained${}^{16}$. The full
linearized $\eta$-function is given by
$$
\rho\eta=\rho s_0-{nk_{{}_B}\over 2}\sum^\infty_{r=1}{1\over r!}a^{(r)}(\vec
r,t) a^{(r)}(\vec r,t)~~,\eqno (15)
$$
so that here the temperature $\Theta$ becomes identical to the
local equilibrium temperature if the moments, not the fluxes, are
defined as the state variables. This deserves some comments. 
It appears that if an exact solution to the Boltzmann equation is used
to compute the function $\rho\eta$ even in the linear approximation,
the phenomenological temperature $\Theta$ is adequately interpreted by
the kinetic temperature $T$. Notice that this would not be the case of
the physical fluxes $\vec q$ and $\cal T$ are taken as variables
instead of the moments $a^{(2)}$ and $a^{(3)}$ to which they are
related. Moreover the higher moments are independent state variables whose
influence on the state of the gas appears through the wave vector
and frequency dependence of the transport coefficients${}^{17-19}$.
\ms
Another point that needs be considered, is the steady state
equation of heat conduction in relation to eq. 11. In the thirteen 
moments approximation 
such an equation, as it has been explicitly shown in ref. 17, is 
identical in form
with Fourier's law in which the local temperature $T$ is involved.
Pursuing the results of ref. 17, the calculations with the 26-moments 
approximation are in very good agreement with the results of
generalized hydrodynamics${}^{20,21}$. These calculations show that
the classical Fourier law is still valid and need not to be replaced by
the non-linear form of eq. (11). The thermal conductivity is wave
vector dependent (see eq. (4.8) and (4.9) in ref. 17) and these
results agree with those obtained by Alder et al.${}^{21}$ using
numerical methods.
\ms
We may thus conclude that, on the basis of the Grad's moment solution to
the Boltzmann equation, the correct consistency condition between the
the kinetic interpretation temperature through the average kinetic
energy of the molecules and eq. (3) is only achieved if the exact
solution, the infinite series for $f(\vec r, \vec c, t)$,  is used. The
kinetic temperature $T(\vec r, t)$ is related to $\Theta$  as defined
in eq. (3) through equations which depend on the number of moments
raised to the status of state variables. It is then clear that $\Theta$
has no bearing whatsoever on the physical description of the state of
the gas.
\ms
One could now turn this question over to the realm of more microscopic
theories capable with dealing with systems other than the dilute gas.
This has been done for instance by Nettleton for real fluids${}^{22-24}$
where the temperature is associated with the intensive variable
conjugate to the systems energy (Hamiltonian). Then of course the
temperature defined through the average kinetic energy differs from the
one interpreted as the average of the total energy. Although for 
equilibrium states this is inmaterial${}^{4b}$, for non-equilibrium states
the question is
which of the two gives a theoretical value which is in closest agreement
with the one obtained through a thermometer reading. The answer is yet
to be given. It is perhaps appropriate at this stage to remind the
reader that this is an old question. Already in 1966${}^{25,26}$
discussions where offered to point out the difference between two
non-equilibrium temperatures defined as it is mentioned above. As it
turns out, these definitions are related with the evaluation of the bulk
viscosity. The value of this coefficient is different depending on
which interpretation of temperature is adopted. Clearly the most
appropriate one would be that which leads to a bulk viscosity that is
in better agreement with experiment. Unfortunately this is a difficult
quantity to measure, but some progress along these lines has been made
for dense fluids ${}^{27}$.
\ms
We conclude that the non-linearities present in the quantities
$\eta$ and $\Theta$ indeed reflect themselves in
measurable properties of the system such as its transport coefficients,
but they have no effect whatsoever in the steady state expressions for
the non conserved variables, let them be the fluxes or in general the
moments of the distribution function $f$ for dilute gases. This
same fact seems to hold for other systems except that the details 
of this influence is much harder to exhibit, although some hints have 
been given using more sophisticated
methods of statistical mechanics.${}^{24,28-34}$. 
\ms
In our opinion Meixner's conjecture seems to prevail for
nonequilibrium processes beyond local equilibrium. One may define
several functions $\eta$ and consequently will obtain an equal
number of equations of state for $\Theta$. Nevertheless, the only
non-equilibrium temperature which has a clear physical meaning and is
consistent with the laws of thermodynamics is $T(\vec r,t)$.
Therefore, no quantity $\Theta$ defined or introduced into a
non-equilibrium theory through eq. (3) has any bearing whatsoever with
the concept of temperature.

\bs
\noi {\bf References.}
\sm
\item {1)} S.R. de Groot and P. Mazur; {\sl Non-equilibrium
Thermodynamics} (Dover Publications Mineola, N.Y. 1984).
\sm
\item {2)} L.S. \gc ~and F.J. Uribe; J. Non-equilib. Thermodyn
\ub{16}, 89 (1991). 
\sm
\item {3)} H.B. Callen, {\sl Thermodynamics} (John Wiley and Sons, New
York  1st. ed., 1960). 
\sm
\item {4)} a) B.C. Eu; Phys. Rev. E \ub{51}, 768 (1995); b) B.C. Eu and
L.S. Garc\'\i a-Col\'\i n (to be published).
\sm
\item {5)} J. Meixner; Rheol. Acta \ub{12}, 465, (1973). See also this
author's contribution in {\sl Foundations of Continuum thermodynamics}
ed. by J.J. Delgado, M.R. Nina and J.W. Whitelaw (Mac Millan, London
1974) p. 129.
\sm
\item {6)} H. Grad; Comm. Pure Appl. Math. \ub{14}, 323 (1961).
\sm
\item {7)} The literature on this particular point is too broad to cite
every work. A recent book by M.C. Mackey, {\sl Time's Arrow: The
Origins of Thermodynamic Behavior} (Springer-Verlag. Berlin, 1992)
provides a good review. See also papers by J.L. del R\y o and L.S. \gc ,
Phys. Rev. E \ub{48}, 819 (1993) and references therein, and in Physica
A \ub{209}, 385 (1994).
\sm
\item {8)} a) D. Jou and J. Casas V\'azquez, Phys. Rev. A. \ub{45}, 
8371 (1992); b) ibid E  \ub{49}, 1040 (1994); c) D. Jou, J.
Casas-V\'azquez and G. Lebon, {\sl Extended Irreversible
Thermodynamics} (Springer-Verlag, Berlin, 1993). 
\sm
\item {9)} I. M\"uller; Zeits. f\"ur Physik \ub{198}, 329 (1967).
\sm
\item {10)} a) W. Hoover, B.L. Holian, H.A. Posch Phys. Rev. E \ub{48}, 
3196 (1993); b) K. Henyes, ibid \ub {48}, 3199 (1994);c) D. Jou and J.
Casas-V\'azquez, ibid \ub{48}, 3201 (1994).
\sm
\item {11)} See any text on the kinetic theory of gases. The standard
work is the book by S. Chapman and T.G. Cowling, {\sl The Mathematical
Theory of Non-uniform Gases} (Cambridge University Press, Cambridge,
3rd. ed. 1970).
\sm
\item {12)} H. Grad, {\sl Principles of the Kinetic Theory of Gases} in
Handbuch der Physik, S. Fl\"ugge, ed. (Springer-Verlag, Berlin 1958)
Vol. XII p. 205.
\sm
\item {13)} S. Harris; {\sl An Introduction to the Theory of the
Boltzmann Equation} (Holt, Rinehart and Winston, Inc. New York, 1971)
Chap. 7.
\sm
\item {14)} L.S. \gc ~and G. Fuentes-Mart\y nez;J. Stat. Phys.\ub{29},
392 (1982).
\sm
\item {15)} R.M. Velasco and L.S. \gc ; J. Stat. Phys.\ub{69}, 217 (1992). 
\sm
\item {16)} R.M. Velasco and L.S. \gc ; J. Non-equilib. Thermodyn.
\ub{19}, 157 (1993). 
\sm
\item {17)} R.M. Velasco and L.S. \gc ; Phys. Rev. A \ub{44}, 4961, (1991). 
\sm
\item {18)} T. Ruggieri and I. Mueller; {\sl Extended Thermodynamics}
(Springer-Verlag, Berlin 1993).
\sm
\item {19)} R.M. Velasco and L.S. \gc ; J. Non-equilib. Thermodyn.
\ub{20},  1, (1995).
\sm
\item {20)} S. Yip and J.P. Boon; {\sl Molecular Hydrodynamics} (Dover
Publications, Mineola, N.Y. 1991).
\sm
\item {21)} W.E. Alley and B. J. Alder, Phys. Rev. A \ub{27}, 3158
(1983); W.E. Alley, B. J. Alder and S. Yip, ibid \ub{27}, 3174 (1983). 
\sm
\item {22)} R. E. Nettleton; Phys. Rev. A\ub{42}, 4622 (1990); Ann. der
Physik \ub{2}, 490 (1993)..
\sm
\item {23)} R. E. Nettleton; J. Phys. A, Math. Gen. \ub{22}, 5281 (1989);
J. Chem. Phys. \ub{93}, 8247 (1990); S. Afr. J. Phys. \ub{14}, 27  (1991).
\sm
\item {24)} R. E. Nettleton; J. Phys. Soc. Japan \ub{61}, 3103 (1992).
J. Chem. Phys. \ub{99}, 3059 (1993).
\sm
\item {25)} L.S. \gc~ and M.S. Green; Phys. Rev. \ub{150}, 153 (1966).
\sm
\item {26)} M. H. Ernst, Physica \ub{32}, 252 (1966).
\sm
\item {27)} H. van Beijeren, J. Karkheck and J.V. Sengers; Phys. Rev.
\ub{A37}, 2247 (1988).
\sm
\item {28)} L.S. \gc , A. Vasconcellos and R. Luzzi; J. Non-equilib. 
Thermodyn. \ub{19}, 24 (1994).  
\sm
\item {29)} A.R. Vasconcellos, R. Luzzi and L.S. \gc , Phys. Rev. A 
\ub {43}, 6822 (1991).
\sm
\item {30)} A.R. Vasconcellos, R. Luzzi and L.S. \gc , Phys. Rev. A 
\ub {43}, 6633 (1991).
\sm
\item {31)} A. R. Vasconcellos, R. Luzzi and L.S. \gc , 
J. Non-Equilib. Thermodyn. \ub{20}, 103 (1995); ibid \ub{20}, 119 (1995).
\sm
\item {32)} J.G. Ramos, A.R. Vasconcellos and R. Luzzi; Fortschr. der
Physik \ub{43}, 266 (1995).
\sm
\item {33)} J. Keizer; J. Chem. Phys. \ub{82}, 2751 (1985) and
references therein. 
\sm
\item {34)} J. Keizer; {\sl Statistical Thermodynamics of Irreversible
processes} (Springer-Verlag, Berlin 1987).
\bye